\documentclass{article}
\usepackage{spconf,amsmath,graphicx, amssymb}
\usepackage{bm}
\usepackage{multirow}
\usepackage{booktabs}
\usepackage{tikz}
\usepackage{float}
\usepackage{hyperref}
\usepackage{arydshln}

\newcommand{\RNum}[1]{\uppercase\expandafter{\romannumeral #1\relax}}

\newcommand{\ra}[1]{\renewcommand{\arraystretch}{#1}}

\title{Speech separation with large-scale self-supervised learning}
\name{\begin{tabular}{c}
Zhuo Chen, 
Naoyuki Kanda,
Jian Wu, 
Yu Wu,
Xiaofei Wang,
Takuya Yoshioka,\\
Jinyu Li,
Sunit Sivasankaran,
Sefik Emre Eskimez
\end{tabular}}
\address{   
Microsoft, One Microsoft Way, Redmond, USA}

\begin{document}
\ninept
\maketitle

\begin{abstract} 

Self-supervised learning (SSL) methods such as WavLM have shown promising speech separation (SS) results in small-scale simulation-based experiments. In this work, we extend the exploration of the SSL-based SS by massively scaling up both the pre-training data (more than 300K hours) and fine-tuning data (10K hours). We also investigate various techniques to efficiently integrate the pre-trained model with the SS network under a limited computation budget, including a low frame rate SSL model training setup and a fine-tuning scheme using only the part of the pre-trained model. Compared with a  supervised baseline and the WavLM-based SS model using feature embeddings obtained with the previously released 94K hours trained WavLM, our proposed model obtains 15.9\% and 11.2\% of relative word error rate (WER) reductions, respectively, for a simulated far-field speech mixture test set. For conversation transcription on real meeting recordings using continuous speech separation, the proposed model achieves 6.8\% and 10.6\% of relative WER reductions over the purely supervised baseline on AMI and ICSI evaluation sets, respectively, while reducing the computational cost by 38\%.
\end{abstract}
\begin{keywords}
WavLM, speech separation, conversation transcription, self-supervised learning, multi-speaker
\end{keywords}

\section{Introduction}
\label{sec:intro}
\vspace{-.505em}

Speech separation (SS) has long been studied as an effective front-end method to handle overlapping speech and quick speaker transitions in conversations. 
Recently, there have been tremendous improvements in the SS model architecture~\cite{Dual_path-Luo2020,hershey2016deep,subakan2021attention,wang2022tf}, enabling their successful applications to conversation transcription~\cite{von2019all,yoshioka2019advances,chen2021conformer,sivaraman2022adapting}.

Most SS models are trained by using overlapping speech signals that are artificially generated from clean speech data rather than using real noisy recordings. It is because the SS training requires clean signals as the training targets.
Such a simulation-based training scheme faces two limitations.
Firstly, 
 although an unlimited amount of overlapped speech data can be generated by simulation, the available clean speech data are limited because they should not include background noise and reverberation to be used as the training targets. This results in inherently limited speech variations even after the simulation~\cite{wisdom2020unsupervised}.
Secondly, the simulated far-field speech mixtures often exhibit a systematic mismatch with real recordings, leading to potentially sub-optimum separation performance on real data~\cite{wang2022leveraging}.

There were several attempts to leverage real noisy data for training SS models.
In \cite{wisdom2020unsupervised,tzinis2022remixit}, a combination-invariant loss function was proposed for training an SS model 
only from noisy audio.
On the other hand, a series of studies on self-supervised learning (SSL) \cite{chen2022wavlm,tsai2022superb,huang2022investigating} 
 showed that the SS quality could be significantly improved 
 by leveraging the SSL pre-trained models. 
Particularly, WavLM \cite{chen2022wavlm} showed prominent improvement in SS, achieving the state-of-the-art performance in the LibriCSS benchmark~\cite{chen2021conformer}.
However, the previous investigations had several limitations. 
Firstly, the amount of the pre-training data 
was limited only to 94K hours, which can be massively extended by using 
real-world recordings.
Secondly, no consideration was made for the inference-time computational cost, where 
a naive application of the SSL pre-trained models often incurs an unacceptable increase in the inference cost.
Thirdly, the past evaluation of SSL-based SS was only limited to simulation data,
and the effectiveness for real recordings was unexplored.

In this work, we explore the SSL-based SS and its deployment in conversation transcription systems further with the dual objective of improving the SS performance and reducing the system's computation complexity.
Our exploration is conducted based on WavLM \cite{chen2022wavlm} as a backbone of the SSL model.
We first examine the impact of the pre-training data size by drastically increasing the pre-training data quantity for more than 300K hours. 
We also investigate various practical techniques to leverage
the pre-trained model under the limited inference computation budget, 
including a low frame rate SSL training setup and a fine-tuning scheme using only the part of the pre-trained model.
The SS network is integrated into a continuous SS (CSS)-based conversation transcription system and evaluated by using 
both simulation and real recordings from the AMI~\cite{carletta2005ami} and ICSI~\cite{janin2003icsi} meeting corpora.

\vspace{-.505em}
\section{Related Works}
\label{sec:relwork}
\vspace{-.505em}

\subsection{WavLM}
\label{sec:wavlm}
\begin{figure}
\centering
  \includegraphics[width=0.3\textwidth]{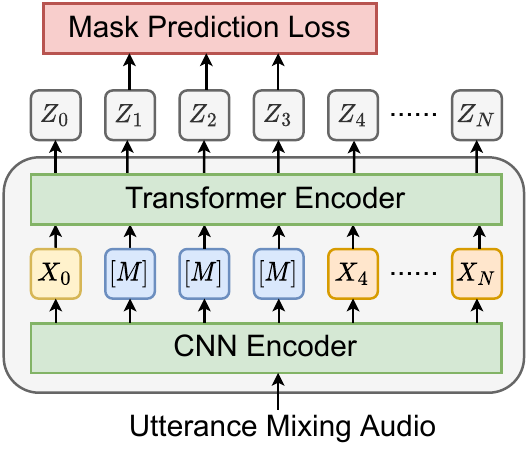}
  \vspace{-5mm}
  \caption{Overview of WavLM. $M$ denotes the masked frames for MSP loss during training }
  \label{fig:wavlm}
  \vspace{-5mm}
\end{figure}

\vspace{-.505em}
WavLM is a self-supervised speech model that achieves the state-of-the-art performance in multiple downstream tasks~\cite{chen2022large,chen2022wavlm,chen2022does}. WavLM follows the learning scheme of masked speech prediction (MSP). 
In this scheme, a pseudo label is first estimated for each frame of unlabeled audio. During training, a random mask is applied to each input audio, and the network is optimized to predict the pseudo label of the masked region, as shown in Fig. \ref{fig:wavlm}.

WavLM distinguishes itself from other MSP-based SSL models by introducing an ``utterance mixing'' procedure, where a training speech sample is randomly mixed with another one.
This data augmentation forces the network to ignore interfering speakers and predict the labels from the main speakers. In this way, speaker discrimination is promoted in WavLM representations, making it beneficial for downstream tasks requiring speaker distinction, such as speaker verification, speaker diarization, and SS. The pseudo label is obtained by clustering unsupervised representations from the unlabeled data.
Following \cite{chen2022does}, we adopt an online speaker clustering tokenizer in WavLM training.

\vspace{-.505em}
\subsection{CSS}
\vspace{-.505em}
CSS is introduced as a front-end to handle overlapping speech
in long-form multi-talker recordings~\cite{yoshioka2018multi,yoshioka2019low,yoshioka2018recognizing}.
In the CSS framework, an input long-form audio is segmented into a set of overlapping chunks.  Each chunk contains frames representing the history, current and future frames, denoted by $T_h$, $T_c$, and $T_f$, respectively.
A local SS network processes each chunk to estimate $N$ short overlap-free signals.  
$N$ long-form overlap-free signals are then obtained by 
stitching the short overlap-free signals (i.e., the SS outputs) from each chunk. 
See \cite{chen2021conformer} for more details.

\vspace{-.505em}
\section{Method}
\vspace{-.505em}


\subsection{WavLM-based SS network}
\vspace{-.505em}

We train an SS network with WavLM by following the training scheme in \cite{chen2022wavlm}.
Specifically, we first train WavLM by using unlabeled speech data. 
We then fine-tune the conformer-based SS network~\cite{chen2021conformer}
by feeding a concatenation of an audio spectrogram and the WavLM embedding as an input.
The WavLM embedding is computed by the weighted average of outputs from all transformer layers of WavLM,
where the weights are learned during the fine-tuning.
We either update or freeze the parameters of WavLM during the fine-tuning.
The SS network is optimized 
by utterance-level permutation invariant training loss~\cite{yu2017permutation} with $N=2$,
where each output branch estimates a magnitude mask of each speaker. 
Fine-tuning is performed on the artificially mixed dataset, whose detail will be introduced in Sec. \ref{sec:fine-tune-config}.
During the inference, the output for each speaker is obtained through a speech masking~\cite{wu2021investigation}.

\vspace{-.505em}
\subsection{Data scale up}
\label{sec:data-scale-up}
\vspace{-.505em}

Three data sets with different data scales are explored for WavLM training. We denote them by small ($S$), medium ($M$), 
and large ($L$) based on  their data sizes.
Data set $S$ is the  same data set used in the original WavLM paper~\cite{chen2022wavlm} and consists of 94k hours of speech. The combination of $S$ and the 220K-hour data from \cite{wang2021unispeech}, totaling 314K hours, serves as data set $M$. 
Data set $L$ is even larger than $M$. 
While we are not disclosing further details of this data set due to company's confidentiality policy, 
we believe that the experimental results with $S$ and $M$ should provide concrete insights regarding the impact of the data sale.  

The training data, except the data $S$, were collected from an in-house dataset from a variety of data sources. 
Because most of the data more or less contain noise, reverberation, and distortion, 
they are not appropriate for the overlapped speech simulation for supervised SS training.
In our preliminary experiments, we observed that a supervised SS model trained with the above-mentioned data showed significantly worse results than an SS model trained with our fine-tuning data generated from the clean signals (detailed later).

\vspace{-.505em}
\subsection{Inference cost reduction}
\vspace{-.505em}

To reduce the inference cost of the WavLM-based SS, we investigate three aspects of model configurations, namely,
balancing the model sizes for WavLM and SS networks, low frame rate WavLM, and a partial usage of WavLM.

\begin{table}[t]
\ra{0.9}
\setlength{\tabcolsep}{2pt}
    \centering
    \footnotesize
    \caption{Model configurations for SSL and SS networks.}
    \label{tab:size}
    \begin{tabular}{cccccc} 
         \toprule 
        SSL model & Params (M) & nLayers & nHeads & nDim & nFFdim\\
         \midrule
         WavLM Small & 53 & 12 &  12 & 384 & 1536\\
         WavLM Base & 90 & 12 & 12 & 768 & 3072\\
         WavLM Large & 300 & 24 & 16 & 1024 & 4096  \\
         \midrule
         SS model & Params (M) & nLayers & nHeads & nDim & nFFdim\\
         \midrule
         SS-9.5 & 9.5 & 8 & 4 & 256 & 1024  \\
         SS-26 & 26 & 16 & 4 & 256 & 1024\\
         SS-59 & 59 & 18 & 8 & 512 & 1024\\
         SS-79 & 79 & 24 & 8 & 512 & 1024\\
         SS-92 & 92 & 28 & 8 & 512 & 1024\\
         \bottomrule
    \end{tabular}
    \vspace{-5mm}
\end{table}

\vspace{-.505em}
\subsubsection{Balancing model sizes for WavLM and SS networks}
\label{sec:model_config}
\vspace{-.505em}

The model size configurations for the SSL pre-trained network and the SS network 
under a limited computation budget
has not been sufficiently explored
for SS.
Generally speaking, a large SSL model offers a better generalization ability for downstream tasks, allowing  
a tiny SS network to be used. 
On the other hand, under the same computation budget, it can be the case where assigning a larger capacity to the SS network by using a small SSL model results in better SS performance. 

Thus, we examine various combinations of WavLM and SS networks with different sizes
as shown in Table \ref{tab:size}.
WavLM consists of a transformer~\cite{vaswani2017attention} with relative position bias~\cite{chi2022xlm}. We examine three WavLM model configurations: Small, Base, and Large.
The SS network consists of a conformer~\cite{chen2021conformer} by following \cite{wu2021investigation}.
We test five model configurations for the SS network 
as shown in Table ~\ref{tab:size}. The model size varies from 9.5 million to 92 million parameters.



\vspace{-.505em}
\subsubsection{Low frame rate WavLM}
\vspace{-.505em}

In the WavLM-based SS framework, different frame rates can be used for the pre-trained and SS models. 
It is often beneficial to use a high frame rate (or equivalently, a small frame shift) for SS modeling~\cite{luo2019conv} as it helps the SS model capture the fine-grained details of the speech. However, this results in an increase in the computational cost.  
Meanwhile, the symbolic representation prediction, as used by WavLM, can be modeled by using a relatively lower frame rate (or a larger frame shift) because the prediction targets do not change too frequently.

To reduce the computational cost needed for the WavLM embedding extraction,
we examine the lower frame rate modeling, i.e., using a larger frame shift for the WavLM pre-trained models. 
In this work, a frame shift $f^{\rm ss}=10$ ms is used for the SS networks, while WavLM employs $f^{\rm wl}=20$ ms as the default frame shift.
We also explore 30-ms and 40-ms frame shifts for WavLM to examine the possibility of further reducing the inference cost. 
During fine-tuning, to align the time sequence between the WavLM embedding and spectrogram, the WavLM embedding is duplicated by $f^{\rm wl}/f^{\rm ss}$ times before concatenation.

\vspace{-.505em}
\subsubsection{Higher WavLM layer removal}
\vspace{-.505em}


It is known that the contribution of each layer of an SSL model is different for each downstream task.
Specifically, it is shown that the lower layers are more beneficial for speaker-related tasks such as SS and speaker diarization, while higher layers are essential for semantic tasks such as speech recognition \cite{chen2022wavlm}. 

Based on this observation, we explore fine-tuning the SS network with the WavLM embeddings that are computed by using only lower layers of the pre-trained models. This can significantly reduce the inference cost.  
We examine various configurations, which will be presented in Section \ref{sec:result_and_discussions}.

\vspace{-.505em}
\section{Experiment settings}
\vspace{-.505em}

\subsection{Pre-training configuration}
\label{sec:pre}
\vspace{-.505em}
%
Various WavLM models as shown in Table \ref{tab:size} 
were trained using the data $S$, $M$ or $L$ (Section \ref{sec:data-scale-up}) by 
following the training recipe in \cite{chen2022wavlm}. 
Specifically, 
we optimized the network based on the masked speech prediction loss \cite{hsu2021hubert}
while adding the noise or interference speaker speech to the input audio \cite{chen2022wavlm}. 
All WavLM models were trained on 32 NVIDIA A100 GPU cards. Adam optimizer with a peak learning rate of $5e^{-5}$ and a polynomial decay of 0.01 was used. 
The large and base model had the same batch size and training steps as in \cite{chen2022wavlm}. We maximized the batch size for the small model to fill the GPU memory and set the training steps to 600K.

\begin{table}[t]

\ra{0.9}
 \setlength{\tabcolsep}{5.0pt}
    \centering
    \footnotesize
    \caption{Utterance-wise evaluation for data size and model configuration variation. During the fine-tuning,  WavLM was frozen.}
    \label{tab:2}
    \begin{tabular}{llcccccc} 
         \toprule 
        ID &
         \multicolumn{2}{c}{SSL} & SS  & RTF & \multicolumn{2}{c}{WER (\%)} \\ \cmidrule{2-3} \cmidrule{6-7}
       &  \multicolumn{1}{c}{Model} & Data &                &     & Far-mix & Clean-mix\\
         \midrule
    B1&     - & - & SS-59      &$\times$ 0.21 & 22.7 & 22.7 \\
    B2&     - & - & SS-79   & $\times$ 0.27 & 23.2 &  23.8\\
    B3&     - & - & SS-92        & $\times$ 0.32 & 23.1 & 23.6\\
         \midrule
    P1&     WavLM Large & S & SS-59   &  $\times$ 0.55  & 21.5 & 22.8\\
    P2&     WavLM Large & M & SS-59   &  $\times$ 0.55  & 20.6 & 18.2\\
    P3&     WavLM Large & L & SS-59   & $\times$ 0.55  & 19.1 & 17.5 \\  \hdashline[1pt/2pt]\hdashline[0pt/1pt]
    P4&     WavLM Large & L & SS-26   & $\times$ 0.47  & 19.2 & 20.0\\
    P5&     WavLM Base & L & SS-26   & $\times$ 0.25 & 20.4 & 19.2\\
    P6&     WavLM Small & L & SS-26   & $\times$ 0.20 & 20.2 & 20.2\\
         \bottomrule
    \end{tabular}
    \vspace{-5mm}
\end{table}

\begin{table}[t]
\ra{0.9}
\setlength{\tabcolsep}{1.5pt}
    \centering
    \footnotesize
    \caption{Utterance-wise evaluation for frame shift variation and partial layer fine-tuning (FT). WavLM was pre-trained by using the data $L$, and frozen during the fine-tuning.}
    \label{tab:3}
    \begin{tabular}{llcccccc} 
         \toprule 
        ID &
         \multicolumn{3}{c}{SSL} & SS 
         & RTF & \multicolumn{2}{c}{WER (\%)} \\ \cmidrule{2-4} \cmidrule{7-8}
         & \multicolumn{1}{c}{Model}     & $f^{\rm wl} (ms)$ & FT-layers              &     &      & Far-mix & Clean-mix\\
         \midrule
    P3 &   &    20 &&  & $\times$ 0.55 & 19.1 & 17.5\\
    L1 &    WavLM-Large   & 30  & 24 & SS-59 & $\times$ 0.46 & 21.9 & 24.8\\
    L2 &  &     40  &&  & $\times$ 0.38 & 22.8 & 25.7\\
         \midrule
    P4 &   &   &   24 && $\times$ 0.47  & 19.2 & 20.0\\
    S1 &   &   &   16 && $\times$ 0.38 & 19.1 & 18.7\\
    S2 &    WavLM-Large   & 20   & 12 & SS-26 & $\times$ 0.35 & 19.9 & 18.4\\
    S3 &  &    &   8 && $\times$  0.31 & 19.7 & 18.6\\
    S4 &  &    &   4 && $\times$  0.27 & 21.0 & 21.3\\
         \bottomrule
    \end{tabular}
     \vspace{-5mm}
\end{table}

\vspace{-.505em}
\subsection{Fine-tuning configuration}
\label{sec:fine-tune-config}
\vspace{-.505em}

The SS network was fine-tuned by using an artificially generated overlapping speech.
We used the procedures in \cite{wu2021investigation} to generate training samples, where four types of mixing patterns are simulated: partially overlapped speech, fully overlapped speech, sequential two-speaker speech, and single-speaker. The source data for simulation is selected from public datasets including LibriSpeech~\cite{panayotov2015librispeech}, common voice~\cite{ardila2019common}, VCTK~\cite{veaux2017cstr} and a Microsoft in-house corpus. 
For each training mixture, two clean utterances from different speakers are sampled and convolved with two artificially simulated room impulse responses using image method~\cite{imagemethod}. 
The reverberated utterances were then mixed with a random offset and a random source energy ratio between $-5\sim5$ dB. 
A background noise sampled from DNS challenge noise dataset~\cite{reddy2020interspeech} was then added to the reverberated mixture audio, with an SNR ranging from 10 dB $\sim$ 30 dB. 
As a result, 10K hours of overlapped speech was created as the fine-tuning data. A development set of 20 hours of speech was also generated from the same procedure, which was used for model selection. 

We fine-tuned various SS networks listed in Table \ref{tab:size}.
The window size and shift for short-time Fourier transform (STFT) were set to 32 ms and 10 ms, respectively.
The SS networks were optimized by the three-stage training as follows.
We first shortly warmed up the SS network by training the model with artificially mixed data from VoxCeleb dataset~\cite{nagrani2017voxceleb,chung2018voxceleb2}.
In this warm-up stage, we used the AdamW optimizer with a linear decay learning rate schedule, where the peak learning rate was set to $1e^{-4}$.
We then further fine-tuned the model by using the 10K-hour data described above. During this stage, the WavLM parameters are frozen.
The AdamW optimizer with a learning rate of $1e^{-3}$ and a linear decay is used for SS network training. Finally, we optionally continued fine-tuning with 10 times lower learning rate by unfreezing the WavLM network.

\vspace{-.505em}
\subsection{Evaluation scheme}
\vspace{-.505em}
Two evaluation schemes were used for assessing the proposed WavLM-based SS models, namely utterance-wise evaluation (Sec. \ref{sec:utt}) and continuous evaluation\ref{sec:css}.


\begin{table}[t]

\ra{0.9}
 \setlength{\tabcolsep}{5.0pt}
    \centering
    \footnotesize
    \caption{Utterance-wise evaluation for WavLM-based SS without and with additional fine-tuning without freezing WavLM. S3 model was used for comparison.}
    \label{tab:unfreeze}
    \begin{tabular}{lccccccc} 
         \toprule 
        ID &
         \multirow{2}{*}{\shortstack[c]{Additional fine-tuning\\by unfreezing WavLM?}} &
         RTF & 
         \multicolumn{2}{c}{WER (\%)} \\ \cmidrule{4-5}
        & &     & Far-mix & Clean-mix\\
         \midrule
       S3 &   no   & $\times$ 0.31  & 19.7 & 18.6 \\
       S3'&   yes  & $\times$ 0.31   & 19.3 & 18.0\\
         \bottomrule
    \end{tabular}
    \vspace{-5mm}
\end{table}

\vspace{-.505em}
\subsubsection{Utterance-wise evaluation}
\label{sec:utt}
\vspace{-.505em}

The utterance-wise evaluation was performed on simulation-based test sets where each test sample was made by mixing two utterances.
In this evaluation, an SS model was applied for each test sample without chunking the input, as is done for the CSS. 

The SS models were evaluated based on the word error rate (WER) and real-time factor (RTF).
For WER calculation, the SS model was first applied for each test sample to generate separated speech signals,
and the ASR 
 was then applied for each separated speech signal.
We computed the WER for all permutations between the hypotheses and references,
and selected the best WER among them.
We used the latency-controlled long short-term memory-based hybrid ASR system 
from~\cite{yoshioka2019advances}. 
RTF was measured to assess the inference cost. For each model, we measured the computation time $T_r$ for 2.4-sec audio 
and RTF was defined as $T_r/2.4$. The measurement was performed using one core and one thread from an INTEL i7-6700 3.4GHz CPU, averaging over 100 independent runs.

Two test sets named ``far-mix'' and ``close-mix'' were created for this evaluation.
The far-mix
test set was created by shifting and adding two far-field single-speaker utterances from real in-house meeting recordings,
including the speech from 188 speakers. 
The overlap ratio was controlled to be 40\%. 
The total word count for the ``far-mix'' set was 164,025.
On the other hand, the close-mix test was similarly created by
shifting and adding two close-microphone utterances from real in-house recordings,
including the speech from 172 speakers.
The total word count for the ``close-mix'' set was 47,206.

\vspace{-.505em}
\subsubsection{Continuous evaluation}
\label{sec:css}
\vspace{-.505em}

The continuous evaluation was performed for real meeting recordings based on the CSS framework.
We first applied the CSS with $T_h=0.7$ sec, $T_c=1.6$ sec and $T_f=0.1$ sec, respectively, by using 
the WavLM-based SS as a local SS network.
A single speaker merger was also applied in the CSS process to alleviate the distortion introduced for single speaker region~\cite{chen2021conformer}. 
We then applied 
 the same ASR used for the utterance-wise evaluation for the CSS-separated speech signals.
We computed speaker-agnostic WER by using the
multiple dimension Levenshtein edit distance calculation
implemented in 
ASCLITE \cite{fiscus2006multiple}.

We used the recordings in the AMI \cite{carletta2005ami} and ICSI \cite{janin2003icsi} meeting corpora for this evaluation.
For the AMI corpus, the first channel of the microphone array signals is used in the experiment. For the ICSI corpus, we used the signal from the D2 microphone. Before sending it to the SS network, a causal
logarithmic-loop-based automatic gain control is applied on both corpora.   
The evaluation was conducted based on the utterance-group segmentation \cite{kanda2021large}, 
where a long-form recording is segmented at
silence positions or non-overlapping utterance boundaries.

\begin{table}[t]
\ra{0.9}
 \setlength{\tabcolsep}{3.5pt}
    \centering
     \footnotesize
    \caption{Utterance-wise evaluation for various models having similar computation cost as SS-59. Models are sorted in the order of RTF. The data $L$ was used for SSL pre-training.}
    \label{tab:4}
    \begin{tabular}{llcccccc} 
         \toprule 
        ID &
         \multicolumn{2}{c}{SSL} & SS &  RTF & \multicolumn{2}{c}{WER (\%)} \\\cmidrule{2-3}  \cmidrule{6-7}
         & \multicolumn{1}{c}{Model} & FT-layers       &               &      & Far-mix & Clean-mix\\
         \midrule
       B1 & - & -&  SS-59  & $\times$ 0.21 & 22.7 &22.7 \\  \hdashline[1pt/2pt]\hdashline[0pt/1pt]
       S3' & WavLM Large & 8  & SS-26  & $\times$ 0.31 & 19.3 & 18.0\\
       S4' & WavLM Base & 12 & SS-26  & $\times$ 0.25 & 19.6 & 18.2\\
       S5' & WavLM Base & 4  & SS-26  & $\times$ 0.21 & 20.9 & 20.0\\
       S6' & WavLM Small & 12 & SS-26  & $\times$ 0.20 & 20.0 & 18.8\\
       S7' & WavLM Small & 8  & SS-26  & $\times$ 0.19 & 20.9 & 20.0\\
       S8' & WavLM Base  &12 & SS-9.5  & $\times$ 0.19 & 20.1 & 21.0\\
       S9' & WavLM Small & 8  & SS-9.5  & $\times$ 0.13 & 20.6 & 18.5\\
         \bottomrule
    \end{tabular}
    \vspace{-5mm}
\end{table}

\begin{table}[t]
\ra{0.9}
 \setlength{\tabcolsep}{1.0pt}
    \centering
    \footnotesize
    \caption{Continuous evaluation for real meeting recordings.  The data $L$ was used for SSL pre-training.}
    \label{tab:css}
    \begin{tabular}{llcccccccc} 
         \toprule 
        ID &
         \multicolumn{2}{c}{SSL} & SS &RTF& \multicolumn{2}{c}{AMI WER (\%)} && \multicolumn{2}{c}{ICSI WER (\%)}\\ \cmidrule{2-3}\cmidrule{6-7} \cmidrule{9-10}
        & \multicolumn{1}{c}{Model} & FT-layers   &   &   & dev & eval && dev & eval\\  \midrule
       B1&  - & - & SS-59 & $\times$ 0.21 & 21.6 & 25.0 &&  23.2 & 20.7   \\ \hdashline[1pt/2pt]\hdashline[0pt/1pt]
       S3'&  WavLM Large &  8  & SS-26 & $\times$ 0.31 &  19.1 & 22.6 &&  17.8 & 16.5   \\
       S4'&  WavLM Base & 12  & SS-26 & $\times$ 0.25 &  19.4 & 22.9  && 18.6 & 17.2  \\
       S8'&  WavLM Base & 12 & SS-9.5 & $\times$ 0.19 &  19.5 & 22.9 && 18.0 & 17.0   \\
       S9'&  WavLM Small & 8  & SS-9.5 & $\times$ 0.13 &  19.6 & 23.3 && 18.3 & 18.5   \\
         \bottomrule
    \end{tabular}
    \vspace{-5mm}
\end{table}

\vspace{-.505em}
\section{Results and discussions}
\label{sec:result_and_discussions}
\vspace{-.505em}

\vspace{-.505em}
\subsection{Utterance-wise evaluation on simulated test sets}
\vspace{-.505em}
\subsubsection{Impacts of data and model sizes}
\vspace{-.505em}
Table \ref{tab:2} shows the results of utterance-wise evaluation for various data and model configurations.
We made the following observations from the results:
\begin{itemize}
\setlength\itemsep{0em}
\item  
All three models without WavLM (B1, B2, B3) showed similar WERs, indicating the saturation in their performances for our fine-tuning data.
\item Significant and consistent WER reductions were observed by incorporating WavLM.
For example, 
SS-59 trained with WavLM Large (P3) achieved 15.9\% and 22.9\% relative WER reductions over the best model without WavLM (B1) 
for far-mix and close-mix test sets, respectively. 
\item Increasing the amount of the pre-training data clearly benefited the ASR accuracy.
The model pre-trained with the data $L$ (P3) showed 11.2\% and 23.2\% WER reductions over the model pre-trained with the data $S$ (P1) for far-mix and close-mix test sets, respectively.
Generally, we observed a large improvement in the close-mix test set compared to the far-mix test set. 
It would be caused because
 our pre-training data mainly contained close microphone recordings.

\item Increasing the model size of WavLM showed a clear advantage on the WERs over WavLM Base or Small,
but with a cost of an increase of RTF (P4, P5, P6).
On the other hand, increasing the model size of the SS network 
showed relatively less improvement on the WER.
For example, the model with SS-26 (P4) showed only 0.1 points worse WER compared to the model with SS-59 (P3) for the far-mix test set.

\end{itemize}

Finally, it should be highlighted that SS-26 with WavLM-small (P6) achieved 11.0\% and 11.0\% WER reductions over the model without WavLM (B1) for far-mix and clean-mix test sets, respectively, while having a slightly smaller RTF.

\vspace{-.505em}
\subsubsection{Inference cost reduction}
\vspace{-.505em}
The utterance-wise evaluation results for using the low frame rate WavLM are shown in the upper half of Table \ref{tab:3}.
We observe that the low frame rate WavLM (L1, L2) showed significantly worse WERs compared to the WavLM
with $f^{\rm wl}=20$ msec (P3).
This result suggests that preserving fine-grained details in audio signals is crucial not only for the SS task, but also for the SSL models.

We then evaluated the WavLM-based SS in which only a lower part of WavLM layers was used.
The results are shown in Table \ref{tab:3}.
Unlike the low frame rate modeling, the usage of the lower part of WavLM showed a promising trade-off between
RTF and WERs.
For example, by using only the bottom 8 layers of WavLM (S3), the computation was reduced by 34\% while the WER was increased only  relative 2.5\%. 
Interestingly, there was no WER improvement from the model using 16 layers (S1) to the model using all 24 layers (P4), suggesting that the representation in the higher layer is not beneficial for the SS task. 

Table \ref{tab:unfreeze} shows the impact of additionally fine-tuning the WavLM-based SS by unfreezing the WavLM parameters.
As shown in the table, we observed noticeable improvements for both far-mix and clean-mix.
Given this result, we used the additional fine-tuning for all the remaining experiments.

In table \ref{tab:4}, we further explored combinations of SSL model and SS networks.
We collected model configurations that yielded similar inference costs as the baseline SS-59.
From the table, we observed that all WavLM-based systems demonstrated considerable WER improvements over the pure-supervised baseline (B1) under a similar or even smaller RTF. Particularly, SS-9.5 trained with 8 layers of WavLM Small (S9') enjoyed 38\% less RTF (0.21 RTF to 0.13 RTF), while achieving 9.3\% and 18.5\% of
relative WER reduction over the B1 system for far-mix and clean-mix test sets, respectively.

\vspace{-.505em}
\subsection{Continuous evaluation on real meeting recordings}
\vspace{-.505em}
Table \ref{tab:css} shows the continuous evaluation for real meeting recordings.
We selected representative configurations obtained in the previous experiment.
Here, all WavLM-based SS models again yielded 
significant WER reduction in all test sets compared to the purely supervised baseline (B1).
It should be highlighted that  
the S9' model achieved 6.8\% and 10.6\% WER reduction over the B1 model for the AMI and ICSI evaluation set, respectively.

\vspace{-.505em}
\section{Conclusions}
\vspace{-.505em}

In this work, we conducted a comprehensive investigation of WavLM-based SS 
by massively scaling up both the pre-training data and fine-tuning data,
and evaluated the model from both the WER and RTF perspectives.
The proposed model finally achieved 6.8\% and  10.6\% of relative WER reductions over the purely supervised baseline on AMI and ICSI evaluation sets, respectively, while reducing the computational cost by 38\%.

\vfill
\pagebreak

\bibliographystyle{IEEEtran}

{\small\bibliography{refs}}

\begin{thebibliography}{10}
\providecommand{\url}[1]{#1}
\csname url@samestyle\endcsname
\providecommand{\newblock}{\relax}
\providecommand{\bibinfo}[2]{#2}
\providecommand{\BIBentrySTDinterwordspacing}{\spaceskip=0pt\relax}
\providecommand{\BIBentryALTinterwordstretchfactor}{4}
\providecommand{\BIBentryALTinterwordspacing}{\spaceskip=\fontdimen2\font plus
\BIBentryALTinterwordstretchfactor\fontdimen3\font minus
  \fontdimen4\font\relax}
\providecommand{\BIBforeignlanguage}[2]{{%
\expandafter\ifx\csname l@#1\endcsname\relax
\typeout{** WARNING: IEEEtran.bst: No hyphenation pattern has been}%
\typeout{** loaded for the language `#1'. Using the pattern for}%
\typeout{** the default language instead.}%
\else
\language=\csname l@#1\endcsname
\fi
#2}}
\providecommand{\BIBdecl}{\relax}
\BIBdecl

\bibitem{Dual_path-Luo2020}
Y.~Luo, Z.~Chen, and T.~Yoshioka, ``Dual-path {RNN}: efficient long sequence
  modeling for time-domain single-channel speech separation,'' in \emph{Proc.
  ICASSP}, 2020, pp. 46--50.

\bibitem{hershey2016deep}
J.~R. Hershey, Z.~Chen, J.~Le~Roux, and S.~Watanabe, ``Deep clustering:
  Discriminative embeddings for segmentation and separation,'' in \emph{Proc.
  ICASSP}, 2016, pp. 31--35.

\bibitem{subakan2021attention}
C.~Subakan, M.~Ravanelli, S.~Cornell, M.~Bronzi \emph{et~al.}, ``Attention is
  all you need in speech separation,'' in \emph{Proc. ICASSP}, 2021, pp.
  21--25.

\bibitem{wang2022tf}
Z.-Q. Wang, S.~Cornell, S.~Choi \emph{et~al.}, ``{TF-GridNet}: Making
  time-frequency domain models great again for monaural speaker separation,''
  \emph{arXiv preprint arXiv:2209.03952}, 2022.

\bibitem{von2019all}
T.~von Neumann, K.~Kinoshita, M.~Delcroix \emph{et~al.}, ``All-neural online
  source separation, counting, and diarization for meeting analysis,'' in
  \emph{Proc. ICASSP}, 2019, pp. 91--95.

\bibitem{yoshioka2019advances}
T.~Yoshioka, I.~Abramovski, C.~Aksoylar, Z.~Chen \emph{et~al.}, ``Advances in
  online audio-visual meeting transcription,'' in \emph{Proc. ASRU}, 2019, pp.
  276--283.

\bibitem{chen2021conformer}
S.~Chen, Y.~Wu, Z.~Chen \emph{et~al.}, ``Continuous speech separation with
  {Conformer},'' in \emph{Proc. ICASSP}, 2021, pp. 5749--5753.

\bibitem{sivaraman2022adapting}
A.~Sivaraman, S.~Wisdom \emph{et~al.}, ``Adapting speech separation to
  real-world meetings using mixture invariant training,'' in \emph{Proc.
  ICASSP}, 2022, pp. 686--690.

\bibitem{wisdom2020unsupervised}
S.~Wisdom, E.~Tzinis \emph{et~al.}, ``Unsupervised sound separation using
  mixture invariant training,'' \emph{Advances in Neural Information Processing
  Systems}, vol.~33, pp. 3846--3857, 2020.

\bibitem{wang2022leveraging}
X.~Wang, D.~Wang, N.~Kanda, S.~E. Eskimez \emph{et~al.}, ``Leveraging real
  conversational data for multi-channel continuous speech separation,''
  \emph{Proc. Interspeech}, 2022.

\bibitem{tzinis2022remixit}
E.~Tzinis, Y.~Adi \emph{et~al.}, ``{RemixIT}: Continual self-training of speech
  enhancement models via bootstrapped remixing,'' \emph{IEEE Journal of
  Selected Topics in Signal Processing}, 2022.

\bibitem{chen2022wavlm}
S.~Chen, C.~Wang, Z.~Chen, Y.~Wu \emph{et~al.}, ``{WavLM}: Large-scale
  self-supervised pre-training for full stack speech processing,'' \emph{IEEE
  Journal of Selected Topics in Signal Processing}, 2022.

\bibitem{tsai2022superb}
H.-S. Tsai, H.-J. Chang \emph{et~al.}, ``{SUPERB-SG}: Enhanced speech
  processing universal performance benchmark for semantic and generative
  capabilities,'' \emph{Proc. ACL}, 2022.

\bibitem{huang2022investigating}
Z.~Huang, S.~Watanabe, S.-w. Yang, P.~Garc{\'\i}a \emph{et~al.},
  ``Investigating self-supervised learning for speech enhancement and
  separation,'' in \emph{Proc. ICASSP}, 2022, pp. 6837--6841.

\bibitem{carletta2005ami}
J.~Carletta, S.~Ashby \emph{et~al.}, ``The {AMI} meeting corpus: A
  pre-announcement,'' in \emph{International workshop on machine learning for
  multimodal interaction}.\hskip 1em plus 0.5em minus 0.4em\relax Springer,
  2005, pp. 28--39.

\bibitem{janin2003icsi}
A.~Janin, D.~Baron, J.~Edwards, D.~Ellis \emph{et~al.}, ``The {ICSI} meeting
  corpus,'' in \emph{Proc. ICASSP}, vol.~1, 2003, pp. I--I.

\bibitem{chen2022large}
Z.~Chen, S.~Chen, Y.~Wu \emph{et~al.}, ``Large-scale self-supervised speech
  representation learning for automatic speaker verification,'' in \emph{Proc.
  ICASSP}, 2022, pp. 6147--6151.

\bibitem{chen2022does}
S.~Chen, Y.~Wu, C.~Wang, S.~Liu \emph{et~al.}, ``Why does self-supervised
  learning for speech recognition benefit speaker recognition?'' \emph{Proc.
  Interspeech}, 2022.

\bibitem{yoshioka2018multi}
T.~Yoshioka, H.~Erdogan, Z.~Chen \emph{et~al.}, ``Multi-microphone neural
  speech separation for far-field multi-talker speech recognition,'' in
  \emph{Proc. ICASSP}, 2018, pp. 5739--5743.

\bibitem{yoshioka2019low}
T.~Yoshioka, Z.~Chen, C.~Liu, X.~Xiao \emph{et~al.}, ``Low-latency
  speaker-independent continuous speech separation,'' in \emph{Proc. ICASSP},
  2019, pp. 6980--6984.

\bibitem{yoshioka2018recognizing}
T.~Yoshioka, H.~Erdogan \emph{et~al.}, ``Recognizing overlapped speech in
  meetings: A multichannel separation approach using neural networks,'' in
  \emph{Proc. Interspeech}, 2018, pp. 3038--3042.

\bibitem{yu2017permutation}
D.~Yu, M.~Kolb{\ae}k, Z.-H. Tan \emph{et~al.}, ``Permutation invariant training
  of deep models for speaker-independent multi-talker speech separation,'' in
  \emph{Proc. ICASSP}, 2017, pp. 241--245.

\bibitem{wu2021investigation}
J.~Wu, Z.~Chen, S.~Chen, Y.~Wu \emph{et~al.}, ``Investigation of practical
  aspects of single channel speech separation for {ASR},'' \emph{Proc.
  Interspeech}, 2021.

\bibitem{wang2021unispeech}
C.~Wang, Y.~Wu, S.~Liu, J.~Li \emph{et~al.}, ``{UniSpeech} at scale: An
  empirical study of pre-training method on large-scale speech recognition
  dataset,'' \emph{arXiv preprint arXiv:2107.05233}, 2021.

\bibitem{vaswani2017attention}
A.~Vaswani, N.~Shazeer, N.~Parmar, J.~Uszkoreit \emph{et~al.}, ``Attention is
  all you need,'' in \emph{Advances in neural information processing systems},
  2017, pp. 5998--6008.

\bibitem{chi2022xlm}
Z.~Chi, S.~Huang, L.~Dong \emph{et~al.}, ``{XLM-E}: Cross-lingual language
  model pre-training via electra,'' in \emph{Proc. ACL}, 2022, pp. 6170--6182.

\bibitem{luo2019conv}
Y.~Luo and N.~Mesgarani, ``{Conv-TasNet}: Surpassing ideal time--frequency
  magnitude masking for speech separation,'' \emph{IEEE/ACM trans. on ASLP},
  vol.~27, no.~8, pp. 1256--1266, 2019.

\bibitem{hsu2021hubert}
W.-N. Hsu, B.~Bolte \emph{et~al.}, ``{HuBERT}: Self-supervised speech
  representation learning by masked prediction of hidden units,''
  \emph{IEEE/ACM Trans. on ASLP}, 2021.

\bibitem{panayotov2015librispeech}
V.~Panayotov, G.~Chen, D.~Povey, and S.~Khudanpur, ``Librispeech: an {ASR}
  corpus based on public domain audio books,'' in \emph{Proc. ICASSP}, 2015,
  pp. 5206--5210.

\bibitem{ardila2019common}
R.~Ardila, M.~Branson, K.~Davis, M.~Kohler \emph{et~al.}, ``{Common Voice}: A
  massively-multilingual speech corpus,'' in \emph{Proc. LREC}, 2020, pp.
  4218--4222.

\bibitem{veaux2017cstr}
C.~Veaux, J.~Yamagishi, K.~MacDonald \emph{et~al.}, ``{CSTR VCTK} corpus:
  English multi-speaker corpus for {CSTR} voice cloning toolkit,''
  \emph{University of Edinburgh. CSTR}, 2017.

\bibitem{imagemethod}
J.~Allen and D.~Berkley, ``Image method for efficiently simulating small-room
  acoustics,'' \emph{JASA}, vol.~65, pp. 943--950, 1979.

\bibitem{reddy2020interspeech}
C.~K. Reddy, V.~Gopal \emph{et~al.}, ``The {INTERSPEECH} 2020 deep noise
  suppression challenge: Datasets, subjective testing framework, and challenge
  results,'' \emph{Proc. Interspeech}, 2020.

\bibitem{nagrani2017voxceleb}
A.~Nagrani, J.~S. Chung, and A.~Zisserman, ``{VoxCeleb}: A large-scale speaker
  identification dataset,'' \emph{Proc. Interspeech}, pp. 2616--2620, 2017.

\bibitem{chung2018voxceleb2}
J.~S. Chung, A.~Nagrani, and A.~Zisserman, ``{VoxCeleb2}: Deep speaker
  recognition,'' \emph{Proc. Interspeech}, pp. 1086--1090, 2018.

\bibitem{fiscus2006multiple}
J.~G. Fiscus, J.~Ajot \emph{et~al.}, ``Multiple dimension {Levenshtein} edit
  distance calculations for evaluating automatic speech recognition systems
  during simultaneous speech,'' in \emph{Proc. LREC}, 2006, pp. 803--808.

\bibitem{kanda2021large}
N.~Kanda, G.~Ye \emph{et~al.}, ``Large-scale pre-training of end-to-end
  multi-talker {ASR} for meeting transcription with single distant
  microphone,'' \emph{Proc. Interspeech}, pp. 3430--3434, 2021.

\end{thebibliography}

\end{document}